\documentclass{article}
\usepackage{amsmath,amssymb}
\usepackage[dvips]{graphicx}

\newtheorem{theorem}{Theorem}

\newtheorem{corollary}{Corollary}

\title{Localization of discrete-time quantum walks on a half line \\ via the CGMV method}
\author
{
{Norio KONNO,$^{1, }$\footnote{E-mail: konno@ynu.ac.jp} \quad Etsuo SEGAWA$^{2, }$\footnote{E-mail: segawa.e.aa@m.titech.ac.jp}}\\
{\scriptsize {}$^{1}$ \textit{Department of Applied Mathematics, Faculty of Engineering,}}\\  
{\scriptsize \textit{Yokohama National University, Hodogaya, Yokohama 240-8501, Japan}}\\
{\scriptsize {}$^{2}$ \textit{Department of Value and Decision Science,}}\\ 
{\scriptsize \textit{Tokyo Institute of Technology, Meguro, Tokyo 152-8552, Japan }}\\
}
\date{}
\begin{document}
\maketitle
\noindent
\begin{small}
\textbf{Abstract. }
We study discrete-time quantum walks on a half line by means of spectral analysis. 
Cantero \textit{et al}. \cite{CGMV} showed that 
the CMV matrix, which gives a recurrence relation for the orthogonal Laurent polynomials on the unit circle \cite{CMV1}, 
expresses the dynamics of the quantum walk. 
Using the CGMV method introduced by them, 
the name is taken from their initials, 
we obtain the spectral measure for the quantum walk. 
As a corollary, we give another proof for localization of the 
quantum walk on homogeneous trees shown by Chisaki \textit{et al}. \cite{CHKS}. 
\end{small}
\section{Introduction}
\label{intro}
The random walk (RW) is one of the most important models to analyze problems in various fields. 
The quantum walk (QW) is a generalization of the RW \cite{Meyer} 
with a hope that the QW plays such a role in the quantum field. 
In fact, it has been shown that 
there are many results on useful applications of quantum speed up algorithms of QWs (see \cite{Ambainis} 
for a nice review) and 
approximations of QWs describing physical processes given by Dirac and Schr\"{o}dinger equations \cite{FKK,Strauch,SK,Kurzynski,CBS}. 

The time scaling order of weak convergence of QW with a space-homogeneous quantum coin is the square of corresponding classical one. 
That is, while a random walker spreads in proportion to the square root of time, 
a quantum walker spreads in proportion to time. 
In general, the shape of time scaled distribution is an inverted bell-shape on a bounded support. 
See for the explicit expressions for the densities, for example, \cite{Konno1,Konno2} (one-dimensional infinite lattice $\mathbb{Z}$), 
\cite{WK} (two-dimensional infinite lattice $\mathbb{Z}^2$), 
and \cite{CHKS} ($\kappa$-homogeneous tree $\mathbb{T}_\kappa$). 
The review over the detailed discussions on the weak convergence of QWs can be seen in \cite{KonnoBook}. 
Another important property of QW is localization which is defined by $\limsup_{t\to\infty}P(X_t=0)>0$ in this paper, 
where $X_t$ is the QW at time $t$ starting from the origin. 
The emergence of localization on $\mathbb{Z}$ is shown in the case of 
three- and four- dimensional quantum coin \cite{IKS,IK} by using the spatial Fourier transform, 
and spatial inhomogeneous quantum coin by using the spatial Fourier transform \cite{LS}, 
and a path counting method \cite{KonnoLocal}. 
Through numerical simulations, localization occurs with a periodic perturbation \cite{W}. 
Localization in an aperiodic perturbation is shown in \cite{ShikanoKatsura} by using a ``self duality" property of the QW. 
Localization on the other graphs have been investigated, for example, $\mathbb{Z}^2$ \cite{WK,IKK}, 
semi-infinite one-dimensional lattice ($\mathbb{Z}_+=\{0,1,2,\dots\}$)\cite{Oka} and $\mathbb{T}_\kappa$ \cite{CHKS} by using the spatial Fourier transform and 
a path counting. 

In this paper, 
we analyze the spectral measure and the corresponding Laurent polynomials of the unitary operator 
describing the one-step time evolution of the whole system of the QW. 
In particular, we focus on localization of the QWs. 
The analysis is based on the CGMV method introduced by Cantero \textit{et al} \cite{CGMV}. 
The name CGMV is taken from the initial of the authors \cite{CGMV}.  
Path counting methods \cite{Konno1,Konno2} or the Fourier analysis \cite{IKS,Grimmett} are useful for the homogeneous case. 
However, in general, computations in these methods become complicated for the inhomogeneous case, for example, 
one defect model \cite{KonnoLocal}. 
On the other hand, the spectral theory on the unit circle has a long history and has been well studied 
(see \cite{CMV1,Sze,BS,BS2,CMV2}). 
The CGMV method used here is based on the theory. 
So it is expected that it would skip some involved procedures in path counting methods and the Fourier analysis. 
Indeed, generating functions of the amplitudes corresponding to the limit measures can be easily obtained by using the 
Carath\'{e}odory functions in the CGMV method, as compared with usual two methods \cite{CHKS,Oka}. 
Let $\mu$ be a probability measure on $\partial \mathbb{D}=\{z\in \mathbb{C}: |z|=1\}$, where $\mathbb{C}$ is the set of complex numbers. 
$L_\mu^2(\partial \mathbb{D})$ denotes the Hilbert space of $\mu$-square-integrable functions 
on $\partial \mathbb{D}$. 
The Laurent polynomials $\{\chi_l(z) \}_{l=0}^{\infty}$ are orthogonal polynomials 
on $\partial \mathbb{D}$ obtained by applying the Gram-Schmidt orthonormalization 
to $\{1,z,z^{-1},z^2,z^{-2},\dots\}$ with respect to the following inner product: 
\[ (f,g)=\int_{z\in \partial\mathbb{D}} \overline{f(z)}g(z)d\mu(z), \]
where $f,g\in L_\mu^2(\partial \mathbb{D})$. 
By the recurrence relation between Szeg\"{o} polynomials coming 
from $\{1,z,z^2,\dots\}$ on $\partial\mathbb{D}$ \cite{Sze} and the relation between 
the Laurent and Szeg\"{o} polynomials, a recurrence relation for the Laurent polynomials 
can be obtained (see \cite{BS,BS2} for the detailed discussion). 
It is shown in \cite{CMV2} that the CMV matrix, which is a five-diagonal band representation associated 
with the Verblunsky parameter $(\alpha_0,\alpha_1,\alpha_2,\dots)$, gives the simplest expression 
for the recurrence relation of the Laurent polynomials. 
The CGMV method gives us the CMV matrix \cite{CMV1} corresponding the QW. 
The Fourier transform \cite{Grimmett} gives us simpler computational method for the QW than the path counting method \cite{Konno1,Konno2} in some cases. 
However since the Fourier transform maps the whole state space of the QW to the quantum coin space, 
it is hard to recover some aspects of the QWs. 
On the other hand, the spectral analysis for the CMV matrix studied in \cite{CGMV,BS,BS2} systematically gives us 
a more efficient computation of the full system of the QW. 
In this paper, we show that by using the CGMV method, the localization of the QW derives 
from the point mass of the spectral measure on $\partial \mathbb{D}$. 
Furthermore, we give a necessary and sufficient condition for 
localization and the stationary distribution discussed in 
\cite{IKS,IKK,SKJ} for some class of QWs using the CGMV method. 
As a corollary, we obtain the stationary distribution of QW on $\mathbb{T}_\kappa$ which was 
shown in \cite{CHKS} using a path counting method and the Fourier transform. 

The rest of this paper is organized as follows. 
We consider the following two cases: 
(1) the CMV matrix with null-odd Verblunsky parameter, \textit{i.e.}, $(\alpha_0,\alpha_1,\alpha_2,\dots)$
$=(\alpha,0,\alpha,0,\dots)$ and 
(2) null-even Verblunsky parameter, \textit{i.e.}, $(\alpha_0,\alpha_1,\alpha_2,\dots)
=(0,\alpha,0,\alpha,\dots)$. 
In Section 2, we explain the relations between the CMV matrix and the corresponding QWs. 
That is, we clarify a relation between the CMV matrix with null-odd Verblunsky parameter 
and the QW on $\mathbb{Z}_+$ with a self loop at the origin (Type I QW), and a relation between 
the CMV matrix with null-even Verblunsky parameter and the QW with a reflection wall studied in \cite{CHKS,Oka}. 
In Section 3, we present results on localization for Types I and II QWs 
with a general spatial constant coin. 
We discuss the relation between 
localizations of both Types of QWs and the eigenvectors 
at the point mass of spectral measure for the corresponding CMV matrix. 
As a consequence, we give another proof for localization of the QW on $\mathbb{T}_\kappa$ 
shown in \cite{CHKS}.
Finally, in Section 4, we summarize on results. 
\section{Quantum walks and CMV matrices}
The CGMV method is a new analytical tool for QWs introduced by Cantero et al. \cite{CGMV} 
The method is different from path counting \cite{CHKS,Konno1,Konno2} 
and the Fourier analysis \cite{CHKS,IKS,IK,Grimmett}. 
We can get all information on QW from the eigen system of the unitary operator $W$ which describes the 
one-step time-evolution for the system. 
Cantero \textit{et al.} \cite{CGMV} have shown that $W$ can be expressed by 
a CMV matrix $\mathcal{C}$ related to Szeg\"{o} polynomials \cite{CMV1,CMV2} 
and analyze the spectral measure of the CMV matrix 
corresponding to $W$. In this section, we define the two types of QWs (called Types I and II QWs) 
and give the corresponding CMV matrix, respectively. 

Let $\mathcal{C}_{(\alpha_0,\alpha_1,\dots)}$ be the CMV matrix associated with the Verblunsky parameter $(\alpha_0,\alpha_1,\dots)$. 
The simplest matrix representation for $\mathcal{C}_{(\alpha_0,\alpha_1,\dots)}$ is given by the following form \cite{CMV1,CMV2}:
\begin{equation}
\mathcal{C}_{(\alpha_0,\alpha_1,\dots)}=
\begin{bmatrix}
\overline{\alpha}_0      & \rho_0\overline{\alpha}_1    & \rho_0\rho_1  & 0               & 0             & 0               & 0            & 0      & \ldots \\
\rho_0         & -\alpha_0\overline{a}_1      & -\alpha_0\rho_1    & 0               & 0             & 0                & 0             & 0       & \ldots \\
0              & \rho_1\overline{\alpha}_2    & -a_1\overline{\alpha}_2 & \rho_2\overline{\alpha}_3 & \rho_2\rho_3  & 0                & 0             & 0       & \ldots \\  
0              & \rho_1\rho_2       & -\alpha_1\rho_2    & -\alpha_2\overline{\alpha}_3   & -\alpha_2\rho_3    & 0               & 0            & 0       & \ldots \\
0              & 0                  & 0             & \rho_3\overline{\alpha}_4 & -\alpha_3\overline{\alpha}_4 & \rho_4\overline{\alpha}_5 & \rho_4\rho_5 & 0       & \ldots \\
0               & 0                   & 0              & \rho_3\rho_4    & -\alpha_3\rho_4    & -\alpha_4\overline{\alpha}_5   & -\alpha_4\rho_5   & 0      & \ldots  \\
               &\vdots              &\vdots         &\vdots           & \vdots        & \vdots          & \ddots       &  &         
\end{bmatrix},
\end{equation}
where $\alpha_j\in \mathbb{C}$ satisfies $|\alpha_j|< 1$ and $\rho_j=\sqrt{1-|\alpha_j|^2}$. 
At first we present the relation between the QW and the CMV matrix. 
We give definitions of Types I and II QWs, respectively. 
Let the so called ``quantum coin'' $U$ be denoted by  
\begin{equation}
U=\begin{bmatrix} c_{RR} & c_{LR} \\ c_{RL} & c_{LL}\end{bmatrix} \in \mathrm{U}(2), 
\end{equation}
where $\mathrm{U}(2)$ is the set of $2 \times 2$ unitary matrices. 
We define $\rho\equiv |c_{RR}|$, $\mathrm{arg}(c_{RR})\equiv \sigma_R$, $\mathrm{arg}(c_{LL})\equiv \sigma_L$, and 
$\mathrm{det}(U)\equiv \Delta$, where $\mathrm{arg}(z)$ is the argument of $z\in \mathbb{C}$. 
By the unitarity of $U$, we have $\rho= |c_{LL}|$, $\Delta=e^{i(\sigma_R+\sigma_L)}$, and $c_{RL}=-\Delta \overline{c_{LR}}$. 
\begin{enumerate}
\item Type I QW \\
The total space $\mathcal{H}^{(I)}$ for Type I QW is generated by a standard basis 
$\{ |0,S\rangle, \\ |0,L\rangle, |1,R\rangle, |1,L\rangle,|2,R\rangle,|2,L\rangle,\dots \}$. 
The time evolution $W^{(I,U)}$ with quantum coin $U$ is denoted by 
\begin{align*}
W^{(I,U)}|0,S\rangle &= c_{RR}|1,R\rangle+c_{LR}|0,S\rangle, \;\;
W^{(I,U)}|0,L\rangle = c_{RL}|1,R\rangle+c_{LL}|0,L\rangle, \\
W^{(I,U)}|x,R\rangle &= c_{RR}|x+1,R\rangle+c_{LR}|x-1,L\rangle \;\; (x\geq 1), \\
W^{(I,U)}|x,L\rangle &= c_{RL}|x+1,R\rangle+c_{LL}|x-1,L\rangle \;\; (x\geq 1).
\end{align*}
\item Type II QW \\
The total space $\mathcal{H}^{(II)}$ for Type II QW is generated by a standard basis \\
$\{ |0,L\rangle, |1,R\rangle, |1,L\rangle,|2,R\rangle,|2,L\rangle,\dots \}$. 
Let $\gamma\in \mathbb{R}$, where $\mathbb{R}$ is the set of real numbers. 
The time evolution $W^{(II,U)}$ is denoted by 
\begin{align*}\notag
W^{(II,U)}|0,L\rangle &= e^{i\gamma}|1,R\rangle, \\
W^{(II,U)}|x,R\rangle &= c_{RR}|x+1,R\rangle+c_{LR}|x-1,L\rangle \;\; (x\geq 1), \\
W^{(II,U)}|x,L\rangle &= c_{RL}|x+1,R\rangle+c_{LL}|x-1,L\rangle \;\; (x\geq 1). 
\end{align*}
\end{enumerate}
In this paper, we restrict initial states of Types I and II QWs to 
$\Psi^{(I)}_0=\alpha|0,S\rangle+\beta|0,L\rangle$ and $\Psi^{(II)}_0=e^{i\delta}|0,L\rangle$, respectively, 
where $|\alpha|^2+|\beta|^2=1$ and $\delta\in \mathbb{R}$. 
Define a total weight of the passage from position $y$ to position $x$ at time $t$ with the quantum coin $U$ by 
\begin{multline*}
\Xi^{(J,U)}_{x,y}(t)\equiv \sum_{d_1,d_2\in\{R,L\}}I_{\{(x,d_1),(y,d_2)\in \mathcal{H}^{(J)}\}}(d_1,d_2) \\
\times 
\left\{ \langle x,d_1|\left(W^{(J,U)}\right)^t| y,d_2\rangle \right\} |d_1\rangle\langle d_2|
\;\;\;(J\in \{I,II\}), 
\end{multline*} 
where $|R\rangle={}^T[1,0]$, and $|L\rangle={}^T[0,1]$ and $I_A(x,y)$ is the indicator function of $A$, that is, 
$I_A(x,y)=1$ $((x,y)\in A)$, $=0$ $((x,y)\notin A)$. 
Here $T$ is the transposed operator. 
Let $X_t^{(I)}$ and $X_t^{(II)}$ be Types I and II QWs, respectively. 
Then the probability that a particle is measured in location $x$ at time $t$ is defined by 
$P(X^{(J,U)}_t=x)=|| \Xi^{(J)}_{x,0} \varphi_0^{(J)} ||^2$ $(J\in \{I,II\})$, 
where $\varphi_0^{(J)}={}^T[\alpha,\beta]$ $(J=I)$, $=e^{i\delta}$ $(J=II)$. 
To get the matrix representation of $W^{(I,U)}$, 
we give a one-to-one correspondence between the basis of $\mathcal{H}^{(I)}$ and $\{0,1,2,\dots\}$, such that 
$(0,S)\leftrightarrow 0$, $(0,L)\leftrightarrow 1$, $(k,R)\leftrightarrow 2k$, $(k,L)\leftrightarrow 2k+1$ ($k\geq 1$). 
Put $\Lambda_I=\mathrm{diag}[1,\lambda_1^{(I)},\lambda_2^{(I)},\dots]$ with $\lambda_{2k}^{(I)}=e^{-ik\sigma_R}$ ($k\geq 0$), 
$\lambda_{2k-1}^{(I)}=e^{ik\sigma_L}$ ($k \geq 1$), where $\mathrm{diag}[\lambda_0,\lambda_1,\dots]$ is a diagonal matrix, that is, 
$\left(\mathrm{diag}[\lambda_0,\lambda_1,\dots]\right)_{l,m}=\delta_{l,m}\lambda_l$. 
Then we see 
\begin{equation}\label{conju1}
W^{(I,U)}= \Lambda_I{}^T\mathcal{C}_{(a_0,0,a_2,0,\dots)} \Lambda_I^*, 
\end{equation}
where $a_j=a\Delta^{-(j+1)/2}$ $(j=even)$, $=0$ $(j=odd)$ with $a= \overline{c_{LR}}\Delta^{1/2}$. 
Equation (\ref{conju1}) gives a relation between the null-odd CMV matrix and Type I QW.
On the other hand, for Type II QW, a one-to-one correspondence between the basis of 
$\mathcal{H}^{(II)}$ and $\{0,1,2,\dots\}$ is given by 
$(0,L)\leftrightarrow 0$, $(k,R)\leftrightarrow 2k-1$, $(k,L)\leftrightarrow 2k$ ($k\geq 1$). 
Define $\Lambda_{II}\equiv \mathrm{diag}[1,\lambda_1^{(II)},\lambda_2^{(II)},\dots]$ 
with $\lambda_{2k}^{(II)}=e^{-ik(\sigma_L-\gamma)}$, $\lambda_{2k+1}^{(II)}=e^{ik(\sigma_R-\gamma)}$. 
Thus we also see 
\begin{equation}\label{conju2}
W^{(II,U)}= e^{i\gamma} \Lambda_{II}\mathcal{C}_{(0,b_1,0,b_3,\dots)}\Lambda_{II}^*, 
\end{equation}
where $b_j=b(e^{-i\gamma}\Delta)^{-(j+1)}$ $(j=odd)$, $=0$ $(j=even)$ with $b=\overline{c_{LR}}\Delta e^{-i\gamma}$. 
Equation (\ref{conju2}) gives a relation between the null-even CMV matrix and Type II QW. 
Denote the Laurent polynomials for $\mathcal{C}_{(\alpha_0,\alpha_1,\dots)}$ 
as $\widehat{u}_j(z)$ $(j=0,1,\dots)$ and $\widehat{v}_j(z)$ satisfying 
$\widehat{\boldsymbol{u}}(z)\mathcal{C}_{(\alpha_0,\alpha_1,\dots)}=z\widehat{\boldsymbol{u}}(z)$ and 
$\mathcal{C}_{(\alpha_0,\alpha_1,\dots)}\widehat{\boldsymbol{v}}(z)=z\widehat{\boldsymbol{v}}(z)$ respectively, 
where 
$\widehat{\boldsymbol{u}}(z)=[\widehat{u}_0(z),\widehat{u}_1(z),\dots ]$ and 
$\widehat{\boldsymbol{v}}(z)={}^T[\widehat{v}_0(z),\widehat{v}_1(z),\dots ]$. 
If the original Verblunsky parameter $(\alpha_0, \alpha_1, \alpha_2,\dots)$ is changed to \\
$(\alpha_0 e^{iw}, \alpha_1 e^{2iw}, \alpha_2 e^{3iw},\dots)$, then 
the corresponding Laurent polynomials and spectral measure can be rewritten as follows: 
\begin{align}
\widehat{u}_{2k-1}(z) &\to e^{ikw}\widehat{u}_{2k-1}(e^{-ikw}z), \;\;
\widehat{u}_{2k}(z)\to e^{-ikw}\widehat{u}_{2k}(e^{-ikw}z), \label{LP2}  \\ 
\widehat{v}_{2k-1}(z) &\to e^{-ikw}\widehat{u}_{2k-1}(e^{-ikw}z), \;\;
\widehat{v}_{2k}(z)\to e^{ikw}\widehat{u}_{2k}(e^{-ikw}z), \label{LP1} \\ 
\mu(z) &\to \mu(e^{-ikw}z). \label{SP}
\end{align}
Let $\alpha\in \mathbb{C}$ with $|\alpha|<1$ and 
\begin{equation*}
C(\alpha)=\begin{bmatrix} \sqrt{1-|\alpha|^2} & -\alpha \\ \bar{\alpha} & \sqrt{1-|\alpha|^2}\end{bmatrix}. 
\end{equation*}
From the definition of the weight of a passage and Eqs. (1) and (2) for $\Delta=1$, $\gamma=0$ case, we should remark that 
\begin{align*}
\Xi^{(I,C(\alpha))}_{x,y}(t) &= 
\begin{bmatrix}
\left(\mathcal{C}^t_{(\alpha,0,\alpha,0,\dots)}\right)_{2y,2x} & \left(\mathcal{C}^t_{(\alpha,0,\alpha,0,\dots)}\right)_{2y+1,2x} \\
\left(\mathcal{C}^t_{(\alpha,0,\alpha,0,\dots)}\right)_{2y,2x+1} & \left(\mathcal{C}^t_{(\alpha,0,\alpha,0,\dots)}\right)_{2y+1,2x+1}
\end{bmatrix}  \; (x,y\geq 0), \\
\Xi^{(II,C(\alpha))}_{x,y}(t) &= 
\begin{bmatrix}
\left(\mathcal{C}^t_{(0,\alpha,0,\alpha,\dots)}\right)_{2x-1,2y-1} & \left(\mathcal{C}^t_{(0,\alpha,0,\alpha,\dots)}\right)_{2x-1,2y} \\
\left(\mathcal{C}^t_{(0,\alpha,0,\alpha,\dots)}\right)_{2x,2y-1} & \left(\mathcal{C}^t_{(0,\alpha,0,\alpha,\dots)}\right)_{2x,2y}
\end{bmatrix} \;(x,y\geq 0), 
\end{align*}
where if $i,j<0$, then we put $\left(\mathcal{C}^t_{(\alpha_0,\alpha_1,\dots)}\right)_{i,j}=0$. 
Thus Eqs. (\ref{conju1})-(\ref{SP}) and orthonormality of the Laurent polynomials give 
the weight of a passage with a general quantum coin $U$, $\Xi^{(J,U)}_{x,y}(t)$, as follows: 
\begin{align}
\Xi^{(I,U)}_{x,y}(t) &= (\Delta^{1/2})^t e^{i(x-y)\phi}\times 
D^*(\phi) \Xi_{x,y}^{(I,C(a))} D(\phi), \label{U1C} \\
\Xi^{(II,U)}_{x,y}(t) &= (\Delta^{1/2})^t e^{i(x-y)\phi}\times 
D^*(\psi)\Xi_{x,y}^{(II,C(b))} D(\psi),\label{U2C} 
\end{align}
where 
$\psi=\sigma_R-\gamma$, $\phi=(\sigma_R-\sigma_L)/2$, $D(\theta)=\mathrm{diag}(e^{i\theta/2},e^{-i\theta/2})$, 
$a=\overline{c_{LR}}\Delta^{1/2}$, and $b=\overline{c_{LR}}\Delta e^{-i\gamma}$ with $\Delta=\mathrm{det}(U)$. 
As a summary of this section, Types I and II QWs with quantum coin $U$ defined by Eq. (1) can be re-expressed by the CMV matrices with the 
Verblunsky parameters $(a,0,a,0,\dots)$ and $(0,b,0,b,\dots)$ under the correspondences 
\begin{equation} a=\overline{c_{LR}}\Delta^{1/2},\;\;b=\overline{c_{LR}}\Delta e^{-i\gamma}, \end{equation}
respectively. 
The explicit expressions for $\Xi_{x,y}^{(I,C(a))}$ and $\Xi_{x,y}^{(II,C(b))}$ can be obtained by the spectral 
analysis for $\mathcal{C}_{(a,0,a,0,\dots)}$ and $\mathcal{C}_{(0,b,0,b,\dots)}$ in the next section. 
\section{Localization and point mass of spectral measure}
In this section, we will consider the relation between the spectral measure of the CMV matrix 
and localization of the corresponding QW. 
Here we define ``localization'' as follows: there exists $\Psi^{(J)}\in \mathcal{H}^{(J)}$ such that 
\[ \limsup_{t\to\infty}|\langle \Psi^{(J)}, \left(W^{(J,U)}\right)^t\Psi_0^{(J)}\rangle |>0,\;\; (J\in\{I,II\}). \]   
As we will show, the stationary distributions for Types I and II QWs are 
described by the Laurent polynomials at the point mass of the corresponding spectral measure. 
Let $\mu^{(I)}$ and $\mu^{(II)}$ be the spectral measures for $\mathcal{C}_{(a,0,a,0,\dots)}$ and $\mathcal{C}_{(0,b,0,b,\dots)}$, respectively. 
Let $\{\widehat{x}_j(z)\}_{j=0}^{\infty}$ and $\{\widehat{\chi}_j(z)\}_{j=0}^{\infty}$ denote the Laurent polynomials of 
$\mu^{(I)}$ and $\mu^{(II)}$ satisfying 
\begin{equation}\label{reccurence}
\mathcal{C}_{(a,0,a,0,\dots)}\widehat{\boldsymbol{x}}(z) = z\widehat{\boldsymbol{x}}(z), \;\;
\widehat{\boldsymbol{\chi}}(z)\mathcal{C}_{(0,b,0,b,\dots)} = z\widehat{\boldsymbol{\chi}}(z),
\end{equation}
where $\widehat{\boldsymbol{x}}(z)={}^T[\widehat{x}_0(z),\widehat{x}_1(z),\widehat{x}_2(z),\dots]$ and 
$\widehat{\boldsymbol{\chi}}(z)=[\widehat{\chi}_0(z),\widehat{\chi}_1(z),\widehat{\chi}_2(z),\dots]$ 
with $\widehat{x}_0(z)=\widehat{\chi}_0(z)=1$, respectively. Therefore 
\begin{align*}
\left(\mathcal{C}_{(a,0,a,0,\dots)}^t\right)_{lm} 
	&= \int_{|z|=1}z^t\widehat{x}_l(z)\overline{\widehat{x}_m}(z)d\mu^{(I)}(z), \\
\left(\mathcal{C}_{(0,b,0,b,\dots)}^t\right)_{lm}
	&= \int_{|z|=1}z^t\overline{\widehat{\chi}_{l}}(z)\widehat{\chi}_m(z)d\mu^{(II)}(z). 
\end{align*}
To get the spectral measure $\mu^{(J)}$, 
we compute the Carath\'{e}odory function $F^{(J)}(z)$ $(J\in^\{R,L\})$
which are given by 
\begin{equation}\label{Carathe}
F^{(I)}(z)=\lim_{j\to\infty}\frac{\widetilde{x}_j(z)}{x_j(z)},\;\;
F^{(II)}(z)=\lim_{j\to\infty}\frac{\widetilde{\chi}_j(z)}{\chi_j(z)}\;\;(|z|<1),
\end{equation}
where $\widetilde{x_j}(z)$ and $\widetilde{\chi}_j(z)$ are the Laurent polynomials 
whose Verblunsky parameters are $-\alpha_j$ when the original ones are $\alpha_j$. 
Put $\mathcal{B}^{(J)}=\{ \theta\in[-\pi,\pi): \lim_{r \uparrow 1}F^{(J)}(re^{i\theta})=\infty \}$ for $J\in\{R,L\}$. 
Then the spectral measure is obtained by 
\begin{equation}\label{measure}
d\mu^{(J)}(e^{i\theta})=w^{(J)}(\theta)\frac{d\theta}{2\pi}+\sum_{\theta_0\in \mathcal{B}^{(J)}}m_0^{(J)}(\theta_0)\delta(\theta-\theta_0)d\theta, 
\end{equation} 
where 
\begin{equation}\label{conti_mass_meas}
w^{(J)}(\theta) = \lim_{r\uparrow 1}\mathrm{Re} \left(F^{(J)}(re^{i\theta})\right), \;\;\;\;
m_0^{(J)}(\theta_0) = \lim_{r\uparrow 1}\frac{1-r}{2}F^{(J)}(re^{i\theta_0}). 
\end{equation}
Here $\mathrm{Re}(z)$ is the real part of $z\in \mathbb{C}$. 
If $w^{(J)}(\theta)$ is $L^1$ integrable, then by the Riemann-Lebesgue lemma and Eqs. (\ref{U1C}) (\ref{U2C}), 
$\Xi_{k,0}^{(J)}(t)$ can be given by, for sufficiently large time step $t$, 
\begin{multline}\label{master1}
\Xi_{k,0}^{(I)}(t) \sim (\Delta^{1/2})^t e^{ik(\sigma_R-\sigma_L)/2} \\ \times 
\sum_{\theta_0\in \mathcal{B}^{(I)}}m_0^{(I)}(\theta_0)
\begin{bmatrix}
\widehat{x}_{2k}(e^{i\theta_0}) & e^{-i(\sigma_R-\sigma_L)/2} \widehat{x}_{2k}(e^{i\theta_0})\overline{\widehat{x}_1}(e^{i\theta_0}) \\
e^{i(\sigma_R-\sigma_L)/2}\widehat{x}_{2k+1}(e^{i\theta_0}) & \widehat{x}_{2k+1}(e^{i\theta_0})\overline{\widehat{x}_1}(e^{i\theta_0})
\end{bmatrix},
\end{multline}

\begin{equation}\label{master2}
\Xi_{k,0}^{(II)}(t) \sim (\Delta^{1/2})^t e^{ik(\sigma_R-\sigma_L)/2}\times 
\sum_{\theta_0\in \mathcal{B}^{(II)}}m_0^{(II)}(\theta_0)
\begin{bmatrix}
e^{-i(\sigma_R-\gamma)}\widehat{\chi}_{2k-1}(e^{i\theta_0})  \\
\widehat{\chi}_{2k}(e^{i\theta_0})
\end{bmatrix},
\end{equation}
where $A(t) \sim B(t)$ means $\lim_{t\to\infty} |(A(t))_{l,m}/(B(t))_{l,m}|\to 1$ for $l,m\in\{1,2\}$. 
Here $A(t)$ and $B(t)$ are $2\times 2$ matrices. 
\section{Main results}
Now we give explicit expressions for the limit measures of Types I and II QWs. 

\begin{theorem}
Let $X_t^{(I)}$ be Type I QW whose the Verblunsky parameter $(a,0,a,0,\dots)$ at time $t$ with the initial coin state ${}^T[\alpha,\beta]$ 
starting from the origin. 
The quantum coin is given by 
\begin{equation*}
U=\begin{bmatrix} c_{RR} & c_{RL} \\ c_{LR} & c_{LL} \end{bmatrix}. 
\end{equation*}
with $a=\overline{c_{LR}}\Delta^{1/2}$. 
Then we have 
\begin{equation}\label{stable1}
\lim_{t\to\infty}P(X_t^{(I)}= x)=\frac{\mathrm{Re}(a)^2}{1-\mathrm{Im}(a)^2}|\alpha e^{i\phi/2}+\beta e^{-i\phi/2}\nu_I(a)|^2(1+\nu_I^2(a))\nu_I^{2x}(a),
\end{equation}
where $\phi=(\sigma_R-\sigma_L)/2$, and 
\[
\nu^{(I)}(a)=\frac{\mathrm{sgn} (\mathrm{Re}(a))}{\rho} \{ \sqrt{1-\mathrm{Im}(a)^2}-|\mathrm{Re}(a)| \}. 
\]
Here $\sigma_{R(L)}=\mathrm{arg}(c_{RR(LL)})$, 
$\Delta=\mathrm{det}(U)$ and $\rho=\sqrt{1-|a|^2}$. 
\end{theorem}

From the Eq. (\ref{stable1}), we obtain the necessary and sufficient condition 
for localization of Type I QW on both the quantum coin and the initial state in the following. 

\begin{corollary}
Localization of Type I QW from the origin with the initial coin state ${}^T[\alpha,\beta]$ occurs if and only if 
\begin{equation*}
\mathrm{Re}(a)\neq 0,\;\;\alpha e^{i\phi/2}+\beta e^{-i\phi/2}\nu_I(a) \neq 0. 
\end{equation*}
\end{corollary}

The first condition depends only on the quantum coin and the second one depends not only on the quantum coin 
but also the initial coin state. \\
Next, we give the limit measure for Type II QW. 

\begin{theorem}
Let $X_t^{(II)}$ be Type II QW whose the Verblunsky parameter (0,b,0,b,\dots) at time $t$ with the initial state $e^{i\delta}$ 
starting from the origin. The quantum coin is given by 
\begin{equation*}
U=\begin{bmatrix} c_{RR} & c_{LR} \\ c_{RL} & c_{LL} \end{bmatrix}
\end{equation*}
with $b=\overline{c_{LR}}\Delta e^{-i\gamma}$. 
The weight of right moving from the origin is $e^{i\gamma}$ $(\gamma\in \mathbb{R})$. 
Then we have 
\begin{multline}\label{stable2}
\lim_{t\to\infty}P(X^{(II)}_t=x) \\ =\frac{1+(-1)^{x+t}}{2} \times 
\begin{cases}
| M(b) |^2  & \text{: $x=0$, } \\
| M(b)|^2 (1+1/\{\nu^{(II)}(b)\}^2) \left\{\nu^{(II)}(b)\right\}^{2x} & \text{; $x>0$,}
\end{cases}
\end{multline}
where $\nu^{(II)}(b) = \rho/|1+b|$ and 
\[ M(b)=\left\{1+\mathrm{sgn}\left(|b|^2+\mathrm{Re}(b)\right)\right\}\left|\frac{|b|^2+\mathrm{Re}(b)}{(1+b)^2}\right|. \]
Here $\Delta=\mathrm{det}(U)$ and $\rho=\sqrt{1-|b|^2}$. 
\end{theorem}

From Eq. (\ref{stable2}), the necessary and sufficient condition of localization of Type II QW 
on the quantum coin can be seen as follows. 
Remark that the following necessary and sufficient condition is independent of the initial state $e^{i\delta}$. 

\begin{corollary}
Define $D\equiv \{(x,y)\in \mathbb{R}^2: x^2+y^2<1\}$. 
Let the Verbulunsky parameter $(0,b,0,b,\dots)$ of Type II QW be $b=x+iy$ with $(x,y)\in D$. 
Then the necessary and sufficient condition of localization of Type II QW is given by 
\begin{equation*}
(x,y)\in \{ (x,y)\in D: (x+1/2)^2+y^2>(1/2)^2  \}. 
\end{equation*}
\end{corollary}

The Type II QW for $\alpha=2/\kappa-1$ and $\gamma=0$ (resp. $\gamma=\pi$) corresponds to the discrete-time QW on $\kappa$-regular tree $\mathbb{T}_\kappa$ with 
the uniform initial qubit $\varphi_0^A$ (Case A) (resp. weighted initial qubit $\varphi_0^B$ (Case B)) at the origin \cite{CHKS}, 
where $\varphi_0^{J}={}^T[1/\sqrt{\kappa},w_\kappa/\sqrt{\kappa},w_\kappa^2/\sqrt{\kappa},\dots, w_\kappa^{\kappa-1}/\sqrt{\kappa}]$ with 
$w_\kappa=1$ ($J=I$), $=e^{2\pi i/\kappa}$ ($J=II$). 
As a consequence, we can give the stationary distribution for the QW on $\kappa$-regular tree corresponding to the result shown in \cite{CHKS} as follows. 

\begin{corollary}
Let $Y_t^{\kappa}$ be the distance from the origin of the QW on $\mathbb{T}_\kappa$ at time $t$. 
Then we have 
\begin{equation*}
\lim_{t\to\infty}P(Y_t^{\kappa}=x)=\frac{1+(-1)^{t+x}}{2} \times 
\begin{cases}
C(\kappa)  & \text{: $x=0$, } \\
\kappa C(\kappa)\left(\frac{1}{\kappa-1}\right)^x & \text{; $x>0$,}
\end{cases}
\end{equation*}
where 
\[
C(\kappa)= \begin{cases} 0 & \text{; Case (A), }\\ \left(\frac{\kappa-2}{\kappa-1}\right)^2 & \text{; Case (B).}\end{cases}
\]
\end{corollary}

\noindent{{\it Proofs of Theorems 1 and 2:}} 
We show explicit expressions for the spectral measures and corresponding Laurent polynomials for Types I and II QWs 
in Appendices A and B, respectively. 
Then substituting Eqs. (\ref{LaurentI}), (\ref{point_mass}) into Eq. (\ref{master1}) and Eqs. (\ref{LaurentII}), (\ref{mass2}) into Eq. (\ref{master2}), 
implies the desired conclusion with respect to localization of Types I and II QWs, respectively. $\square$

\section{Summary}
We briefly summarize our results. In this paper we investigated localization and the limit distribution for Types I and II QWs 
by using the CGMV method introduced by Cantero et al. \cite{CGMV}. 
We explicitly computed the spectral measure of the CMV matrix corresponding to the QW. From the point mass of the measure, 
we showed localization of the QW. Furthermore, we obtained the necessary and sufficient condition of localization with 
respect to the quantum coin and the initial state. As a corollary, we gave another proof for localization of the QW on 
homogeneous tree shown in \cite{CHKS}. In addition, we presented the limit distribution for the quantum walk. \\
\noindent \\
\noindent {\bf Acknowledgements} \\
This work 
was partially supported by the Grant-in-Aid for Scientific Research (C) of Japan Society for the 
Promotion of Science (Grant No. 21540118). 
%
%
%
%

\noindent \\
\noindent{\large{\bf Appendix A}} \\
\noindent \\
\noindent We give an explicit expression of spectral measure and corresponding Laurent polynomials for Type I QW. 
The Laurent polynomials $\{x_j(z)\}_{j=0}^{\infty}$ for $W^{(I,U)}$ are given by the following 
recurrence relation (\ref{reccurence}): 
\begin{equation}\label{LaurentI}
\widehat{x}_0(z)=1, \;\;
\widehat{x}_{2n-1}(z)=B_+(z)\lambda_+^n(z)+B_-(z)\lambda_-^n(z),\;\;\widehat{x}_{2n}(z)=\overline{\widehat{x}_{2n-1}}(1/\overline{z}),
\end{equation}
where
\[
B_{\pm}(z)=\frac{\pm(z^{-1}-a)/\rho\mp\lambda_\mp(z)}{\lambda_+(z)-\lambda_-(z)},
\]
with 
\[ \lambda_{\pm}(z)=\frac{1}{2\rho}\left\{ z+z^{-1}\mp \mathrm{sgn}(\mathrm{Im}(e^{-i|\arccos{\rho}|}z))
\sqrt{(z-z^{-1})^2+4|a|^2} \right\}. \]
Here $\mathrm{Im}(z)$ is the imaginary part of $z\in \mathbb{C}$ 
and $\mathrm{sgn}(x)=1$ $(x>0)$, $=0$ $(x=0)$, $-1$ $(x<0)$ for $x\in \mathbb{R}$. 
Then we can compute the Carath\'{e}odory function by Eqs. (\ref{Carathe}) and (\ref{LaurentI}) as follows: 
\begin{equation}\label{CaratheI}
F^{(I)}(z)=\lim_{j\to\infty}\frac{\widetilde{\widehat{x}}_j(z)}{\widehat{x}_j(z)}
	=-\frac{z-z^{-1}-2i\mathrm{Im}(a)}{\sqrt{(z-z^{-1})^2+4|a^2|}-2\mathrm{Re}(a)}. 
\end{equation}
Therefore from Eqs. (\ref{measure}) and (\ref{CaratheI}), we obtain 
\[
w(\theta)=\frac{\sqrt{\rho^2-\cos^2\theta}}{|\sin \theta+\mathrm{Im}(a)|}I_{\{\theta\in [-\pi,\pi): |\cos\theta|<\rho\}}(\theta). 
\]
Furthermore from Eq. (\ref{conti_mass_meas}), we can calculate $\mathcal{B}^{(I)}$ and $m_0^{(I)}$ as follows: 
\begin{align}\label{point_mass}
\mathcal{B}^{(I)} = \{\arcsin (-\mathrm{Im}(a))\},\;\;\;\;
m_0^{(I)}(\theta_0) = \frac{|\mathrm{Re}(a)|}{\sqrt{1-\mathrm{Im}^2(a)}}\;\; (\theta_0\in \mathcal{B}^{(I)}). 
\end{align}
These expressions agree with the ones given in \cite{CGMV}. \\

\noindent \\
\noindent{\large{\bf Appendix B}} \\
\noindent \\
\noindent We give an explicit expression of spectral measure and corresponding Laurent polynomials for Type II QW. 
From the reccurence relation (\ref{reccurence}), we see that 
the Laurent polynomials $\{\chi_j(z)\}_{j=0}^{\infty}$ for $W^{(II,U)}$ are described as 
\begin{align}\label{LaurentII}
\chi_0(z) &=1, \\
\chi_{2n}(z) &=B_+^{(e)}(z)\lambda_+^n(z)+B_-^{(e)}(z)\lambda_-^n(z),\\
\chi_{2n+1}(z) &=B_+^{(o)}(z)\lambda_+^n(z)+B_-^{(o)}(z)\lambda_-^n(z), 
\end{align}
where
\begin{align*}
B_{\pm}^{(e)}(z) = \frac{\pm(z^{-1}-bz)/\rho\mp\lambda_\mp(z)}{\lambda_+(z)-\lambda_-(z)}, \;\;
B_{\pm}^{(o)}(z) = \frac{\pm(z-\overline{b}z^{-1})/\rho\mp\lambda_\mp(z)}{\lambda_+(z)-\lambda_-(z)}\times z. 
\end{align*}
Then we compute the Carath\'{e}odory function by Eqs. (\ref{Carathe}) and (\ref{LaurentII}) as follows: 
\begin{equation}\label{CaratheII}
F^{(II)}(z)
=-\frac{(1-b)z-(1-\overline{b})z^{-1}}{bz+\overline{b}z^{-1}-\sqrt{(z-z^{-1})^2+4|b^2|}}.
\end{equation}
Therefore from Eqs. (\ref{measure}) and (\ref{CaratheII}) we obtain 
\begin{equation*}
w(\theta)
=\frac{\sqrt{|\rho|^2-\cos^2\theta}}{|\mathrm{Im}[(b+1)e^{i\theta}]|}I_{\{\theta\in [-\pi,\pi): |\cos\theta|<\rho\}}(\theta),
\end{equation*}
Furthermore by Eq. (\ref{conti_mass_meas}), 
\begin{align}
\mathcal{B}^{(II)} &= \left\{\theta_0\in[-\pi,\pi):\mathrm{Im}\left((b+1)e^{i\theta}\right)=0\right\}, \notag \\
m_0^{(II)}(\theta_0) &= M(b)/2 \;\;\;\;(\theta_0\in \mathcal{B}^{(II)}), \label{mass2}
\end{align}
where 
\begin{equation} \label{M(b)}
M(b)=\left\{1+\mathrm{sgn}\left(|b|^2+\mathrm{Re}(b)\right)\right\}\left|\frac{|b|^2+\mathrm{Re}(b)}{(1+b)^2}\right|. 
\end{equation}
In particular, in the case of $b=\pm(-1+2/\kappa)$ which corresponds to the QW on $\kappa$-regular tree \cite{CHKS}, we obtain
\begin{multline}
d\mu(\theta)=
(1-C(\kappa))
 \frac{\sqrt{4(\kappa-1)/\kappa^2-\cos^2\theta}}{2/\kappa \cdot \sin \theta}
 I_{\{ \theta\in[-\pi,\pi):|\cos\theta|<2\sqrt{\kappa-1}/\kappa \}}(\theta) \frac{d\theta}{2\pi}\\
+ C(\kappa) \frac{ \delta(\theta)+\delta(\theta-\pi) }{2}d\theta, 
\end{multline}
where $C(\kappa)=0$ if $a=-1+2/\kappa$, $=(\kappa-2)/(\kappa-1)$ if $a=1-2/\kappa$.

\end{document}